\documentclass[aps,pra,amsfonts,amssymb,nofootinbib,twocolumn,groupedaddress]{revtex4}
\usepackage{latexsym}
\usepackage{epsfig} 
\usepackage[utf8]{inputenc}
\usepackage[english]{babel}
\usepackage{amsmath}
\usepackage{color}

\usepackage{esint}
\usepackage{amsthm}

\begin{document}

\title{Achieving the Holevo bound via sequential  measurements}

\author{V. Giovannetti}
\affiliation {NEST, Scuola Normale Superiore and Istituto Nanoscienze-CNR, Piazza dei Cavalieri 7, I-56126 Pisa, Italy}
\author{S. Lloyd} 
\affiliation{Research Laboratory of Electronics, Massachusetts Institute of Technology, Cambridge, MA, 02139, USA}
\author {L. Maccone}
\affiliation{Research Laboratory of Electronics, Massachusetts Institute of Technology, Cambridge, MA, 02139, USA}

\begin{abstract}
  We present a new decoding procedure to transmit classical
  information in a quantum channel which, saturating
  asymptotically the Holevo bound, achieves the optimal  rate of the communication line. Differently from previous
  proposals, it is based on performing a sequence of (projective)
  YES/NO measurements which in $N$ steps determines which codeword was
  sent by the sender ($N$ being the number of the codewords). Our
  analysis shows that as long as $N$ is below the limit imposed by the
  Holevo bound the error probability can be sent to zero
  asymptotically in the length of the codewords.
\end{abstract}

\maketitle

\section{Introduction}

By constraining the amount of
classical information which can be reliably encoded into a collection
of quantum states~\cite{BOUND}, the Holevo bound sets a limit on the
rates that can be achieved when transferring classical messages in a
quantum communication channel. Even though, for finite number of
channel uses, the bound in general is not achievable, it is
saturated~\cite{HOL,SCHU} in the asymptotic limit of infinitely many
channel uses. Consequently, via proper optimization and
regularization~\cite{HASTINGS}, it provides the quantum analog of the
Shannon capacity formula~\cite{COVER}, i.e.  the {\em classical
  capacity} of the quantum channel (e.g. see
Refs.~\cite{BENSHOR,HOLEVOREV}).
 
Starting from the seminal works of Ref.~\cite{HOL,SCHU}  several alternative versions of the asymptotic attainability of the Holevo bound have been  presented so far 
 (e.g. see  Refs.~\cite{HOLEVOREV,WINTER,OGAWA,HAYA,HAYA1} and references therein). 
The original proof~\cite{HOL,SCHU}  was obtained extending to the
quantum regime the typical subspace encoding argument of Shannon communication theory~\cite{COVER}. In this context an
explicit detection scheme  (sometime presented as  the {\em pretty good measurement} (PGM) scheme~\cite{SCHU,HAUS})  was introduced that  allows for exact message recovery
in the asymptotic  limit infinitely long codewords.  More recently, Ogawa and Nagaoka~\cite{OGAWA}, and 
Hayashi and  Nagaoka~\cite{HAYA}  proved the asymptotic  achievability  of the bound by establishing a formal connection with 
quantum hypothesis testing problem~\cite{QHT}, and by generalizing a technique (the information-spectrum method) which was introduced by Verd\'{u} and Han~\cite{VERDU}  in the context of classical communication channel.

In this paper we analyze a new decoding procedure for classical
communication in a quantum channel. Here we give a formal proof using
conventional methods, whereas in \cite{PAPER} we give a more intuitive
take on the argument. Our decoding procedure allows for a new proof of
the asymptotic attainability of the Holevo bound. As in
Refs.~\cite{HOL,SCHU} it is based on the notion of typical subspace
but it replaces the PGM scheme with a sequential decoding strategy in
which, similarly to the quantum hypothesis testing approach of
Ref.~\cite{OGAWA}, the received quantum codeword undergoes to a
sequence of simple YES/NO projective measurements which try to
determine which among all possible inputs my have originated it. To
prove that this strategy attains the bound we compute its associated
{\em average} error probability and show that it converges to zero in
the asymptotic limit of long codewords (the average being performed
over the codewords of a given code {\em and} over all the possible
codes).  The main advantage of our scheme resides on the fact that,
differently from PGM and its
variants~\cite{TYSON,MOCHO,BEL,BAN,UTS,ELD,HAUS,BARN,MONT,FIU,HAYDEN,KHOLEVO},
it allows for a simple intuitive description, it clarifies the role of
entanglement in the decoding procedure \cite{PAPER}, its analysis
avoids some technicalities, and it appears to be more suited for
practical implementations.

 The paper is organized as follows: in Sec.~\ref{INT} we set the problem and present the scheme in an informal, non technical way.   The formal derivation of the
 procedure begins in the next section. Specifically, the notation and some basic definitions  are presented  in Sec.~\ref{sec1}. Next the new sequential detection strategy is
 formalized in Sec.~\ref{sec2}, and finally the main result is derived in Sec.~\ref{sec3}. Conclusions and perspectives are given in Sec.~\ref{seccon}. The paper includes also
 some technical Appendixes. 

\section{Intuitive description of the model}\label{INT}

The transmission of classical messages through a quantum channel can
be decomposed in three logically distinct stages: the {\em encoding}
stage in which the sender of the message (say, Alice) maps the
classical information she wish to communicate into the states of some
quantum objects (the quantum information carriers of the system); the
{\em transmission} stage in which the carriers propagate along the
communication line reaching the receiver (say, Bob); and the {\em decoding}
stage in which Bob performs some quantum measurement on the carriers
in order to retrieve Alice's messages.  For explanatory purposes we
will restrict the analysis to the simplest scenario where Alice is
bound to use only unentangled signals and where the noise in the
channel is memoryless\footnote{\label{NOTA}A similar formulation of the problem
  holds also when entangled signals are allowed: in this case however
  the $\sigma_j$ defined in the text represents (possibly entangled)
  states of $m$-longs {\em blocks} of carriers: for each possible
  choice of $m$, and for each possible coding/decoding strategy one
  define the error probability as in Eq.~(\ref{ff1prova}). The optimal
  transmission rate (i.e.~the capacity of the channel) is also
  expressible as in the rhs term of Eq.~(\ref{capa}) via proper
  regularization over $m$ (this is a consequence of the
  super-additivity of the Holevo information~\cite{HASTINGS}). Finally
  the same construction can be applied also in the case of quantum
  communication channels with memory, e.g. see Ref.~\cite{KRE}.}.
Under this hypothesis the coding stage can be described as a process
in which Alice encodes $N$ classical messages into factorized states
of $n$ quantum carriers, producing a collection ${\cal C}$ of $N$
quantum codewords of the form $\sigma_{\vec{j}} := \sigma_{j_1}\otimes
\cdots \otimes \sigma_{j_n}$ where $j_1,\cdots, j_n$ are symbols
 extracted from a classical alphabet and where we use $N$ different
 vectors $\vec j$.  Due to the
communication noise, these strings will be received as the factorized
states $\rho_{\vec{j}}:=\rho_{j_1} \otimes \cdots \otimes \rho_{j_n}$
(the output codewords of the system), where for each $j$ we have
\begin{eqnarray}
\rho_j = T(\sigma_j) \label{channel} \;,
\end{eqnarray}
$T$ being the completely positive, trace preserving
channel~\cite{NICH} that defines the noise acting on each carrier.
Finally, the decoding stage of the process can be characterized by
assigning a specific Positive Valued Operator Measurement
(POVM)~\cite{NICH} which Bob applies to $\rho_{\vec{j}}$ to get a
(hopefully faithful) estimation $\vec{j}'$ of the value $\vec{j}$.
Indicating with $\{ X_{\vec{j}}, X_0=\openone-\sum_{\vec{j}\in{\cal
    C}} X_{\vec{j}} \}$ the elements which compose the selected POVM,
the average error probability that Bob will mistake a given $\vec{j}$
sent by Alice for a different message, can now be expressed as, e.g.
see Ref.~\cite{HOL},
\begin{eqnarray}\label{ff1prova}
P_{err} := \frac{1}{N} \sum_{\vec{j}\in{\cal C}} (1 - \mbox{Tr}  [ X_{\vec{j}} \rho_{\vec{j}} ])\;.
\end{eqnarray}
In the limit infinitely long sequences $n \rightarrow \infty$, it is known~\cite{HOL,SCHU, HOLEVOREV,WINTER,OGAWA,HAYA} that $P_{err}$ can be  sent to zero under the
 condition that $N$ scales as $2^{n R}$ with $R$ being bounded by  the optimized version of the Holevo information, i.e. 
\begin{eqnarray} \label{capa}
R \leqslant \max_{\{p_j, \sigma_j\}} \; \chi(\{ p_j, \rho_j \}) \;,
\end{eqnarray} 
where the maximization is performed over all possible choices of the inputs  $\sigma_j$ and over all possible probabilities $p_j$, and
where  for a given quantum  output ensemble $\{ p_j, \rho_j\}$  we have
\begin{eqnarray}
\chi(\{ p_j, \rho_j\}) :=  S(\sum_j p_j \rho_j) - \sum_j p_j S(\rho_j)\;,
\end{eqnarray} 
with $S(\cdot) := -\mbox{Tr} [(\cdot) \log_2 (\cdot)]$ being the von Neumann entropy~\cite{NICH}. The inequality in Eq.~(\ref{capa}) is a direct consequence of the Holevo bound~\cite{BOUND}, and its right-hand-side   defines  the so called Holevo capacity of the channel $T$, i.e.   the highest achievable rate of the communication line which guarantees asymptotically null
zero error probability under the constraint of employing only
unentangled codewords\footnote{See footnote \ref{NOTA}.}. 
 In  Refs.~\cite{HOL,SCHU} the achievability of the bound~(\ref{capa}) was obtained by showing that 
 that from {\em any} output quantum ensemble $\{ p_j, \rho_j\}$ it is possible to identify a set of $\sim 2^{n \chi(\{ p_j, \rho_j\}) }$ output codewords $\rho_{\vec{j}}$,
and a decoding POVM 
for which the error probability of Eq.~(\ref{ff1prova}) goes to zero as $n$ increases. Note that proceeding this way, one can forget about the initial mapping
$\vec{j} \rightarrow \sigma_{\vec{j}}$  and work directly with the $\vec{j} \rightarrow \rho_{\vec{j}}$ mapping. This is an important simplification which  typically
is not sufficiently stressed (see however Ref.~\cite{HAYA}).
Within this framework, the proof \cite{HOL,SCHU} exploited the random coding trick  by Shannon  in which the POVM is shown to provide exponential small error probability {\em in average}, when
mediating over all possible groups of codewords associated with $\{ p_j, \rho_j\}$.

\begin{figure}[t!]
\centering
\includegraphics[width=0.48\textwidth,angle=0]{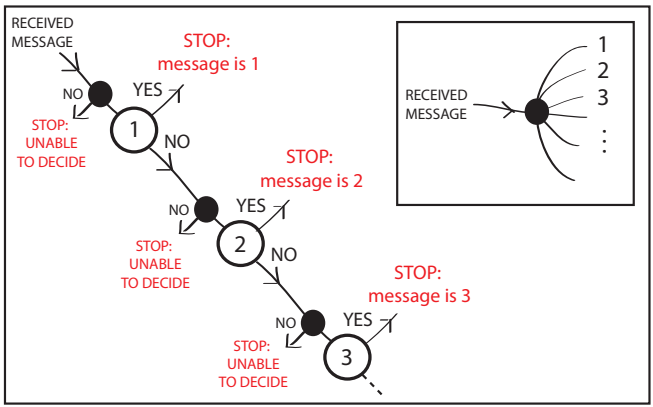}
\caption{Flowchart representation of the detection scheme:  
the projections on the typical subspace ${\cal H}_{typ}^{(n)}(\vec{j})$ of the codewords are represented by the open circles, while the
projections on the typical subspace ${\cal H}_{typ}^{(n)}$ of the average message of the source are represented by the black circles.  The inset describes
 the standard PGM decoding scheme which produces all the possible outcomes in a single step.} \label{fig0}
 \end{figure}

The idea we present here follows the same typicality approach of
Refs.~\cite{HOL,SCHU} but assumes a different detection scheme.  In
particular, while in Refs.~\cite{HOL,SCHU} the POVM produces all
possible outcomes
in a single step as shown schematically in the inset of Fig.~\ref{fig0}, our
scheme is sequential. 
  Namely, Bob performs a
sequence of measurements to test for each of the codewords.
Specifically, he performs a first YES/NO measure to verify whether or
not the received signal corresponds to the first element of the list,
see Fig.~\ref{fig0}.  If the answer is YES he stops and declares
that the received message was the first one. If the answer is NO he
takes the state which emerges from the measurement apparatus and
performs a new YES/NO measure aimed to verify whether or not it
corresponds to the second elements of the list, and so on until he has
checked for all possible outcomes.  The difficulty resides in the fact
that, due to the quantum nature of the codewords, at each step of the
protocol the received message is partially modified by the measurement
(a problem which will {\em not} occur in a purely classical
communication scenario). This implies for instance that the state that
is subject to the second measurement is not equal to what Bob received
from  the quantum channel. As a consequence, to avoid that
the accumulated errors diverge as the detection proceeds, the YES/NO
measurements needs to be carefully designed to have little impact on
the received codewords. As will be clarified in the following section
we tackle this problem by resorting on the notion of typical
subspaces~\cite{SCHUMI}: specifically our YES/NO measurements will be
mild modifications of von Neumann projections on the typical subspaces
of the codewords, in which their non exact orthogonality is smoothed
away by rescaling them through further projection on the typical
subspace of the source average message (see Sec.~\ref{sec2} for
details).

 \section{Sources, Codes and typical subspaces} \label{sec1} In this
 section we review some basic notions and introduce the definitions
 necessary to formalize our detection scheme.

 An independent, identically distributed quantum source is defined by
 assigning the quantum ensemble ${\cal E} = \{ p_j, \rho_j: j\in {\cal
   A} \}$ which specifies the density matrices $\rho_j \in
 \mathfrak{S}({\cal H})$ emitted by the source as they emerge from the
 memoryless channel, as well as the probabilities $p_j$ associated
 with those events (here $j$ is the associated classical random
 variable which takes values on the domain ${\cal A}$).  Since the
 channel is memoryless, when operated $n$ consecutive times, it
 generates products states $\rho_{\vec{j }} \in \mathfrak{S}({\cal
   H}^{\otimes n})$ of the form
\begin{eqnarray}
\rho_{\vec{j}} :=  \rho_{j_1}\otimes  \cdots \otimes \rho_{j_n}\;,\label{eq1}
\end{eqnarray} 
with probability
\begin{eqnarray}
p_{\vec{j}}:=  p_{j_1} p_{j_2} \cdots p_{j_n}\;,\label{rpob1}
\end{eqnarray}
(in these expressions $\vec{j}:= (j_1, \cdots, j_n) \in {\cal A}^n$).
In strict analogy  to  Shannon information theory, 
one defines a $N$-element CODE {\bf C}
as a collection of $N$ states of the form~(\ref{eq1}), i.e. 
\begin{eqnarray} 
\mbox{\bf C} := \{ \rho_{\vec{j}}\in \mathfrak{S}({\cal H}^{\otimes n}) : \vec{j} \in {\cal C}\}\;,
\end{eqnarray} 
with ${\cal C}$ being the subset of $ {\cal A}^n$ which identifies  the elements of  {\bf C} (i.e. the codewords of the code). 
The probability that the source will generate the code {\bf C} can then be  computed as the (joint) probability of emitting all  the codewords that compose it, i.e. 
\begin{eqnarray}\label{ll}
P(\mbox{\bf C}) := \prod_{\vec{j}\in {\cal C}} p_{\vec{j}} = \prod_{\vec{j}\in{\cal C}} \prod_{\ell=1}^n p_{j_\ell} \;.
\end{eqnarray}

\subsection{Typical spaces} 
 
Consider $\rho= \sum_{j} p_j \rho_j \in \mathfrak{S}({\cal H})$ the {\em average} density matrix associated with the ensemble ${\cal E}$, and
let $\rho = \sum_\ell q_\ell |e_\ell\rangle \langle e_\ell|$ its spectral decomposition (i.e. $|e_\ell\rangle$
are the orthonormal basis of ${\cal H}$ formed by the eigenvectors of $\rho$ while $q_\ell$ are their eigenvalues).
For fixed $\delta >0$, one defines~\cite{SCHUMI} the typical subspace ${\cal H}_{typ}^{(n)}$ of $\rho$ as the subspace of ${\cal H}^{\otimes n}$ spanned by those
vectors 
\begin{eqnarray}\label{ffd}
|e_{\vec{\ell}}\rangle := |e_{\ell_1}\rangle \otimes 
\cdots \otimes |e_{\ell_n}\rangle \;, 
\end{eqnarray}
whose associated probabilities $q_{\vec{\ell}} := q_{\ell_1} q_{\ell_2} \cdots q_{\ell_n}$ satisfy the constraint,
\begin{eqnarray}\label{sat}
2^{-n (S(\rho) +\delta)} \leqslant q_{\vec{\ell}}  \leqslant 2^{-n (S(\rho) -\delta)}\;,
\end{eqnarray} 
where $S(\rho) =- \mbox{Tr} [\rho \log_2 \rho]$ is the von Neumann entropy of~$\rho$ 
(as in the classical case~\cite{SHANNON}, the states
  $|e_{\vec{\ell}}\rangle$ defined above can be thought as those which, in average,  contain the symbol $|e_{\ell}\rangle$ almost $n q_\ell$ times).
Identifying with ${\cal L}$ the set of those vectors $\vec{\ell}=(\ell_1,\ell_2, \cdots, \ell_n)$ which satisfies Eq.~(\ref{sat}), 
 the projector $P$ on ${\cal H}_{typ}^{(n)}$ can then be expressed as
 \begin{eqnarray}\label{proj}
P= \sum_{\vec{\ell}\in {\cal L}} \; | e_{\vec{\ell}}\rangle  \langle e_{\vec{\ell}}|\;,
\end{eqnarray} 
while the average state $\rho^{\otimes n}$ is clearly given by 
 \begin{eqnarray}
\rho^{\otimes n}= \sum_{\vec{\ell}} \; q_{\vec{\ell}} \; | e_{\vec{\ell}}\rangle  \langle e_{\vec{\ell}}|\;.
\end{eqnarray}  
By construction, the two operators satisfy the inequalities
\begin{eqnarray} \label{use}
 P \;  2^{-n( S(\rho) + \delta)} \leqslant  P \rho^{\otimes n} P  \leqslant   P \;  2^{-n( S(\rho) - \delta)} \;.
 \end{eqnarray}
 Furthermore, it is known that the probability that ${\cal E}$ will
 emit a message which is not in ${\cal H}_{typ}^{(n)}$ is
 exponentially depressed~\cite{SCHUMI}. More precisely, for all
 $\epsilon>0$ it is possible to identify a sufficiently large  $n_0$
 such for  all $n\geqslant n_0$ we have
 \begin{eqnarray}
  \mbox{Tr} [\rho^{\otimes n}(\openone - P) ]  < \epsilon \;. \label{bbhd1}
  \end{eqnarray}

Typical subsets can be defined also for each of the product states of
Eq.~(\ref{eq1}), associated to each codeword at the output of the channel. In this case the definition is as follows~\cite{HOL}: first for each $j \in{\cal A}$ we define the
spectral decomposition of the element $\rho_j$, i.e.
\begin{eqnarray}
\rho_j = \sum_{k} \lambda_k^j |e_k^{j}\rangle\langle e_k^{j}| \;,
\end{eqnarray} 
where  $|e_k^j\rangle$ are the eigenvectors of $\rho_j$ and $\lambda_{k}^j$ the corresponding eigenvalues (notice that while $\langle e_k^j | e_{k'}^j\rangle = 
\delta_{kk'}$ for all $k,k'$ and $j$, in general the quantities $\langle e_k^j | e_{k'}^{j'}\rangle$ 
are a-priori undefined). Now the spectral decomposition of the codeword $\rho_{\vec{j}}$ is provided by,
\begin{eqnarray}
\rho_{\vec{j}} = \sum_{\vec{k}}  \lambda_{\vec{k}}^{(\vec{j})} \; | e_{\vec{k}}^{(\vec{j})}\rangle \langle e_{\vec{k}}^{(\vec{j})}|, \label{eq2}
\end{eqnarray} 
where for $\vec{k}:= (k_1,\cdots, k_n)$ one has
\begin{eqnarray}
| e_{\vec{k}}^{(\vec{j})}\rangle &:=& |e_{k_1}^{j_1}  \rangle \otimes |e_{k_2}^{j_2}  \rangle \otimes \cdots \otimes |e_{k_n}^{j_n} \rangle
\;,\nonumber \\
\lambda_{\vec{k}}^{(\vec{j})} &:=& \lambda_{k_1}^{j_1}  \lambda_{k_2}^{j_2} \cdots  \lambda_{k_n}^{j_n} \;.
\end{eqnarray} 
Notice that for fixed $\vec{j}$ the vectors $| e_{\vec{k}}^{(\vec{j})}\rangle$ are an orthonormal set  of ${\cal H}^{\otimes n}$; notice also that 
in general such vectors have nothing to do with the vectors $|e_{\vec{\ell}}\rangle$ of Eq.~(\ref{ffd}).

Now the typical subspace ${\cal H}_{typ}^{(n)}(\vec{j})$  of $\rho_{\vec{j}}$ is defined as the linear subspace of ${\cal H}^{\otimes n}$ spanned by the $| e_{\vec{k}}^{(\vec{j})}\rangle$ whose
associated $\lambda_{\vec{k}}^{(\vec{j})}$ satisfy the inequality,
\begin{eqnarray}\label{def1}
2^{-n (S(\rho) -\chi({\cal E})+\delta)} \leqslant  \lambda_{\vec{k}}^{(\vec{j})}  \leqslant 2^{-n (S(\rho) -\chi({\cal E})-\delta)}\;,
\end{eqnarray} 
with 
\begin{eqnarray}
\chi({\cal E}):= S(\rho) -\sum_j p_j S(\rho_j)\;,
\end{eqnarray}
being the Holevo information of the source ${\cal E}$. The projector on 
${\cal H}_{typ}^{(n)}(\vec{j})$  can then be written as
\begin{eqnarray}\label{gh}
P_{\vec{j}} :=  \sum_{\vec{k}\in {\cal K}_{\vec{j}}} \; | e_{\vec{k}}^{(\vec{j})}\rangle  \langle e_{\vec{k}}^{(\vec{j})}|\;,
\end{eqnarray} 
where  ${\cal K}_{\vec{j}}$ identify  the set of the labels   $\vec{k}$ which satisfy Eq.~(\ref{def1}).

We notice that the bounds for the probabilities
$\lambda_{\vec{k}}^{(\vec{j})}$ do not depend on the value of
$\vec{j}$ which defines the selected codeword: they are only function
of the source ${\cal E}$ only (this of course does not imply that the
subspace ${\cal H}_{typ}^{(n)}(\vec{j})$ will not depend on
$\vec{j}$).  It is also worth stressing that since the vectors $|
e_{\vec{k}}^{(\vec{j})}\rangle$ in general are not orthogonal with
respect to the label $\vec{j}$, there will be a certain overlap
between the subspaces ${\cal H}_{typ}^{(n)}(\vec{j})$.  The reason why
they are defined as detailed above stems from the fact that the
probability that $\rho_{\vec{j}}$ will not be found in ${\cal
  H}_{typ}^{(n)}(\vec{j})$ (averaged over all possible realization of
$\rho_{\vec{j}}$), can be made arbitrarily small by increasing $n$,
e.g. see Ref.~\cite{HOL}. More precisely, for fixed $\delta >0$, one
can show that for all $\epsilon >0$ there exists $n_0$ such that for
all $n>n_0$ integer one has,
\begin{eqnarray} \label{IMPO}
\sum_{\vec{j}} \; p_{\vec{j}} \; \mbox{Tr} [ \rho_{\vec{j}} (\openone - P_{\vec{j}})] < \epsilon\;,
\end{eqnarray} 
where $p_{\vec{j}}$ is the probability~(\ref{rpob1}) that the source ${\cal E}$ has emitted the codeword $\rho_{\vec{j}}$.

\subsection{Decoding and Shannon's averaging trick} 

The goal in the design of a decoding stage is to identify a POVM
attached to the code {\bf C} that yields a vanishing error probability
as $n$ increases in identifying the codewords. How can one prove that
such a POVM exists?  First of all let us remind that a POVM is a
collection of positive operators $\{ X_{\vec{j}},
X_0=\openone-\sum_{\vec{j}\in{\cal C}} X_{\vec{j}} : \vec{j} \in {\cal
  C} \}$. The probability of getting a certain outcome $\vec{j}'$ when
measuring the codeword $\rho_{\vec{j}}$ is computed as the expectation
value $\mbox{Tr} [ X_{\vec{j}'} \rho_{\vec{j}} ]$ (the outcome
associated with $\mbox{Tr} [ X_{0} \rho_{\vec{j}} ]$ corresponds to
the case in which the POVM is not able to identify any of the possible
codewords).  Then, the error probability (averaged over all
possible codewords of {\bf C}) is given by the quantity
\begin{eqnarray}\label{ff1}
P_{err}({\mbox{C}}) := \frac{1}{N} \sum_{\vec{j}\in{\cal C}} (1 - \mbox{Tr}  [ X_{\vec{j}} \rho_{\vec{j}} ])\;.
\end{eqnarray}
Proving that this quantity is asymptotically null will be in general quite complicated. However,  the situation simplifies if one averages $P_{err}({\mbox{C}})$
with all codewords {\bf C} that the source  ${\cal E}$ can generate, i.e. 
 \begin{eqnarray}\label{ff12}
\langle{P_{err}}\rangle := \sum_{\mbox{\bf C}} P(\mbox{\bf C}) \; P_{err}({\mbox{C}}) \;,
\end{eqnarray}
$P(\mbox{\bf C})$ being the probability defined in
Eq.~(\ref{ll}). Proving that $\langle{P_{err}}\rangle$
nullifies for $n\rightarrow \infty$ implies that at least one of the
codes {\bf C} generated by ${\cal C}$ allows for asymptotic null error
probability with the selected POVM (indeed the result is even stronger
as {\em almost all} those which are randomly generated by ${\cal C}$
will do the job). In Refs.~\cite{SCHU,HOL} the achievability of the
Holevo bound was proven adopting the pretty good measurement detection
scheme, i.e. the POVM of elements
\begin{eqnarray}
X_{\vec{j}}& =&\Big[ \sum_{\vec{h}\in {\cal C}} P P_{\vec{h}} P \Big]^{-\tfrac{1}{2}} \;  P P_{\vec{j}} P\; \Big[ \sum_{\vec{h}\in{\cal C}} P P_{\vec{h}} P \Big]^{-\tfrac{1}{2}}, \\
X_0 &=& \openone - \sum_{\vec{j}\in{\cal C}} X_{\vec{j}}\;,
\end{eqnarray}
where $P$ is the projector (\ref{proj}) on the typical subspace of the average state of the source, for $\vec{j}\in \mbox{\bf C}$ the $P_{\vec{j}}$ are the projectors~(\ref{gh}) associated
with the codeword $\rho_{\vec{j}}$. 
With this choice one can verify that, for given $\epsilon$ there exist $n$ sufficiently large such that  Eq.~(\ref{ff12}) yields the inequality~\cite{HOL}
 \begin{eqnarray}\label{ff12666}
\langle{P_{err}}\rangle 
\leqslant 4 \epsilon + (N-1) \; 2^{-n (\chi({\cal E})-2 \delta)}  \;.
\end{eqnarray}
This implies that as long as $N-1$ is smaller than $2^{-n (\chi({\cal E})-2 \delta)}$  one can bound the (average) error probability close to zero.

\section{The sequential detection scheme}\label{sec2}

In this section we formalize our detection scheme.

As anticipated in the introduction, the   idea is to determine the value of the label $\vec{j}$ associated with the received codeword $\rho_{\vec{j}}$,  by checking 
whether or not such state  pertains to the typical subspace of the
element $\vec j$ of the selected code ${\cal C}$. 

Specifically we proceed as follows 
\begin{itemize}
\item  first we fix an ordering of the codewords of ${\cal C}$ yielding the sequence $\vec{j}_1, \vec{j}_2, \vec{j}_3, \cdots, \vec{j}_N$ with $\vec{j}_u\in{\cal C}$ for all $u=1, \cdots, N$ 
(this is not really relevant but it is useful to formalize the protocol);
\item   then Bob performs a YES/NO measurement that determines whether or not
the received state is the typical subspace of the first codeword $\vec{j}_1$\footnote{It is worth stressing that in Ref.~\cite{PAPER} this test was implemented by performing a series
of rank-one projective measurements on to a basis of the subspace.}; 
\item if the answer is YES the protocol stops and Bob declares to have identified the received message as the first of the list (i.e. $\vec{j}_1$);
\item  if the answer is NO  Bob, performs a YES/NO measurement to check whether or not
the state is in the typical sub of $\vec{j}_2$;
\item the protocol goes on, testing similarly for all $N$
  possibilities. In the end we will either determine an estimate of
  the transmitted $\vec{j}$ or we will get a null result (the messages
  has not been identified, corresponding to an error in the
  communication).
\end{itemize} 
We now better specify the YES/NO measurements. Indeed, as mentioned
earlier, we have to ``smooth'' them to account for the disturbance
they might introduce in the process.  For this purpose, each of such
measurements will consist in two steps in which first we check (via a von
Neumann projective measurement) whether or not the incoming state is in
the typical subspace ${\cal H}_{typ}^{(n)}$ of the average message.
Then we apply a von Neumann projective measurement on the typical
subspace ${\cal H}_{typ}^{(n)}(\vec{j_i})$ of the $i$-th codeword of
Bob's list (see Fig.~\ref{fig0}).  Hence, the POVM elements are defined as follows.
The first element $E_1$ tests if the transmitted state is in ${\cal H}_{{typ}}^{(n)}(\vec{j}_1)$, so it is described by the (positive) operator
\begin{eqnarray}
E_1 := \bar{P}_{\vec{j}_1}\;, \label{E1}
\end{eqnarray} 
where for any operator $\Theta$ the symbol $\bar{\Theta}$ stands for 
\begin{eqnarray}
\bar{\Theta} := P \Theta P \label{conv}\;,
\end{eqnarray} 
$P$ being the projector of Eq.~(\ref{proj}).
Similarly the remaining elements can be expressed as follows
\begin{eqnarray}
E_2 &:=&  ( \bar{\openone} - \bar{P}_{\vec{j}_1})  \bar{P}_{\vec{j}_2} ( \bar{\openone} - \bar{P}_{\vec{j}_1}) \;, \label{gg} 
\\ \nonumber 
E_3 &:=&  ( \bar{\openone} - \bar{P}_{\vec{j}_1})  (  \bar{\openone} - \bar{P}_{\vec{j}_2})  \; \bar{P}_{\vec{j}_3}  \; ( \bar{\openone}  - \bar{P}_{\vec{j}_2}) (  \bar{\openone}  - \bar{P}_{\vec{j}_1}) \;,\\
\nonumber 
E_4 &:=& \cdots \;,
\end{eqnarray} 
(see Appendix~\ref{APOVM} for an explicit derivation).  
A compact expression can be derived by writing 
\begin{eqnarray}
E_u =M_u^\dag M_u\;, \label{ffd3} 
\end{eqnarray}
where 
\begin{eqnarray} 
M_{u} := {P}_{\vec{j}_u}  P \; \;    \bar{Q}_{\vec{j}_{u-1}}\;  \bar{Q}_{\vec{j}_{u-2} }\; \cdots  \; \bar{Q}_{\vec{j}_{1} }\;,
\end{eqnarray}
with $Q_{\vec{j}}$ being the orthogonal complement of ${P}_{\vec{j}}$, i.e.
\begin{eqnarray}
Q_{\vec{j}} :=  \openone -  {P}_{\vec{j}}  \;. \label{oche1}
\end{eqnarray}
With such definitions the associated average error probability~(\ref{ff12}) can then be expressed as, 
\begin{eqnarray}\label{ff10}
\langle{P_{err}}\rangle &=&
 \sum_{\vec{j}_1, \cdots,  \vec{j}_N } \frac{p_{\vec{j}_1}  \cdots p_{\vec{j}_N} }{N} \sum_{u=1}^N (1 - \mbox{Tr}  [ M_{u} \rho_{\vec{j}_u} M_u^\dag])\nonumber \\
&=&1 - \frac{1}{N} \; \sum_{\vec{j}}\;  p_{\vec{j}}\; \sum_{\ell=0}^{N-1}\;  \mbox{Tr} [ P_{\vec{j}} \;\; \Phi^\ell(\rho_{\vec{j}})]\;,
\end{eqnarray}
where  we  used the fact that the summations over the various $\vec{j}_i$ are independent.
In writing the above expression we introduced the following super-operator
\begin{eqnarray}
\Phi(\Theta) := \sum_{\vec{j}} p_{\vec{j}} \; \bar{Q}_{\vec{j}} \;  \Theta \;  \bar{Q}_{\vec{j}}\;,
\end{eqnarray} 
which is completely positive and trace decreasing, and we use the notation $\Phi^\ell$ to indicate the $\ell$-fold concatenation of super-operators, e.g.  
$\Phi^{2}(\cdot) = \Phi (\Phi( \cdot))$. 
It is worth noticing that the possibility of expressing $\langle{P_{err}}\rangle$ in term of a single super-operator follows
directly from the average we have performed over all possible codes {\bf C}. 
For future reference we find it useful to cast  Eq.~(\ref{ff10}) in a slightly different form by exploiting the
the definitions of  Eqs.~(\ref{eq2}) and (\ref{gh}). More precisely, we write 
\begin{eqnarray}\label{ff144}
&&1- \langle{P_{err}}\rangle 
= \sum_{\ell=0}^{N-1} \sum_{\vec{j},\vec{j}_1,\cdots, \vec{j}_\ell}\;
\frac{p_{\vec{j}} p_{\vec{j}_1}  \cdots p_{\vec{j}_\ell }}{N} \;
\nonumber\\&&
\qquad\qquad\times\mbox{Tr}
[ P_{\vec{j}} \bar{Q}_{\vec{j}_1} \cdots \bar{Q}_{\vec{j}_\ell} \rho_{\vec{j}}  \bar{Q}_{\vec{j}_\ell} \cdots \bar{Q}_{\vec{j}_1} ] \nonumber \\ 
&&= \sum_{\ell=0}^{N-1} \sum_{\vec{j},\vec{j}_1,\cdots, \vec{j}_\ell}\; 
\sum_{\vec{k}} \sum_{\vec{k}'\in{\cal K}_{\vec{j}}}  \; \lambda_{\vec{k}}^{(\vec{j})}  \; 
\frac{p_{\vec{j}} p_{\vec{j}_1}  \cdots p_{\vec{j}_\ell }}{N} \;
\nonumber\\&&\qquad\qquad\times
\left| \langle e_{\vec{k}'} ^{(\vec{j})} |  \bar{Q}_{\vec{j}_1} \cdots \bar{Q}_{\vec{j}_\ell}   |e_{\vec{k}} ^{(\vec{j})}\rangle\right|^2\;. \label{ff122}
\end{eqnarray}

\section{Bounds on the error probability}\label{sec3}

In this section we derive an upper limit for the error probability~(\ref{ff10}) which will lead us to the prove the achievability of  the Holevo bound.

Specifically, we notice that 
\begin{eqnarray}\nonumber 
&&\sum_{\vec{k}} \sum_{\vec{k}'\in{\cal K}_{\vec{j}}}  \; \lambda_{\vec{k}}^{(\vec{j})}  \; 
\big| \langle e_{\vec{k}'} ^{(\vec{j})} |  \bar{Q}_{\vec{j}_1} \cdots \bar{Q}_{\vec{j}_\ell}   |e_{\vec{k}} ^{(\vec{j})}\rangle\big|^2
\\ 
&&\geqslant   \sum_{\vec{k}\in{\cal K}_{\vec{j}}}  \; \lambda_{\vec{k}}^{(\vec{j})}  \; 
\big| \langle e_{\vec{k}} ^{(\vec{j})} |  \bar{Q}_{\vec{j}_1} \cdots \bar{Q}_{\vec{j}_\ell}   |e_{\vec{k}} ^{(\vec{j})}\rangle\big|^2
\nonumber \\
&&=  \sum_{\vec{k}\in{\cal K}_{\vec{j}}}  \; \lambda_{\vec{k}}^{(\vec{j})}  
\big| \langle e_{\vec{k}} ^{(\vec{j})} |  \bar{Q}_{\vec{j}_1} \cdots \bar{Q}_{\vec{j}_\ell}   |e_{\vec{k}} ^{(\vec{j})}\rangle\big|^2 \; \sum_{\vec{k}} \; \lambda_{\vec{k}}^{(\vec{j})} 
\nonumber \\ 
&&\geqslant 
\big|\sum_{\vec{k}\in{\cal K}_{\vec{j}}}  \;  \lambda_{\vec{k}}^{(\vec{j})}  \;
\langle e_{\vec{k}} ^{(\vec{j})} |  \bar{Q}_{\vec{j}_1} \cdots \bar{Q}_{\vec{j}_\ell}   |e_{\vec{k}} ^{(\vec{j})}\rangle\big|^2\nonumber \\
&&  =
\big| \mbox{Tr} [ P_{\vec{j}} \; \rho_{\vec{j}} \;  P_{\vec{j}} \; \; \bar{Q}_{\vec{j}_1} \cdots \bar{Q}_{\vec{j}_\ell} ] \big|^2\;,
\end{eqnarray}
where the first inequality follows by dropping some positive
terms (those with $\vec{k} \neq \vec{k'}$), 
the first identity simply exploits the fact that the $ \lambda_{\vec{k}'}^{(\vec{j})} $ are normalized probabilities when summing over all $\vec{k}$, 
and the second inequality 
follows 
by applying the Cauchy-Schwarz inequality.
Replacing this into Eq.~(\ref{ff122}) we can write 
\begin{eqnarray}\label{ff144rffd}
&& 1- \langle{P_{err}}\rangle \geqslant \\ \nonumber 
&&  \sum_{\ell=0}^{N-1} \sum_{\vec{j},\vec{j}_1,\cdots, \vec{j}_\ell}\; 
\frac{p_{\vec{j}} p_{\vec{j}_1}  \cdots p_{\vec{j}_\ell }}{N} \;
\left| \mbox{Tr} [ P_{\vec{j}} \; \rho_{\vec{j}} \;  P_{\vec{j}} \; \; \bar{Q}_{\vec{j}_1} \cdots \bar{Q}_{\vec{j}_\ell} ] \right|^2\;.
\end{eqnarray}
This can be further simplified by invoking again the   Cauchy-Schwarz inequality  this time with respect to the summation over the ${\vec{j},\vec{j}_1,\cdots, \vec{j}_\ell}$, i.e. 
\begin{eqnarray}\label{ff144rreeee}
&&\sum_{\vec{j},\vec{j}_1,\cdots, \vec{j}_\ell}\; 
p_{\vec{j}} p_{\vec{j}_1}  \cdots p_{\vec{j}_\ell } \;
\left| \mbox{Tr} [ P_{\vec{j}} \rho_{\vec{j}}  P_{\vec{j}} \; \; \bar{Q}_{\vec{j}_1} \cdots \bar{Q}_{\vec{j}_\ell} ] \right|^2 \;  \nonumber \\
&&\geqslant 
\big| \sum_{\vec{j},\vec{j}_1,\cdots, \vec{j}_\ell}\; 
p_{\vec{j}} p_{\vec{j}_1}  \cdots p_{\vec{j}_\ell } \;
 \mbox{Tr} [ P_{\vec{j}}  \rho_{\vec{j}} P_{\vec{j}} \; \; \bar{Q}_{\vec{j}_1} \cdots \bar{Q}_{\vec{j}_\ell} ]  \big|^2 
 \nonumber  \\ 
 &&\qquad \qquad \qquad = \left(
 \mbox{Tr} [ W_1  \; {\cal Q}^\ell ]  \right)^2 \;, 
 \end{eqnarray}
 where for $q$ integer we defined
 \begin{eqnarray}
 W_q &:=& \sum_{\vec{j}} \; p_{\vec{j}}   \; P_{\vec{j}} \; \rho_{\vec{j}}^q \;P_{\vec{j}} \;, \\
{\cal Q} &:=&  \sum_{\vec{j}} \; p_{\vec{j}}  \; \bar{Q}_{\vec{j}_\ell} = \bar{\openone} - \bar{W}_0 \;,
\end{eqnarray} 
(notice that $W_0$ {\em  is not}  $\rho^{\otimes n}$, e.g. see Eq.~(\ref{prima})). 
Therefore one gets 
\begin{eqnarray}\label{ff144rffdd}
1- \langle{P_{err}}\rangle \geqslant  \frac{1}{N} \; \sum_{\ell=0}^{N-1} \left| 
 \mbox{Tr} [ W_1  \; {\cal Q}^\ell ]  \right|^2 \;.
 \end{eqnarray} 
 To proceed it is important to notice  that  the quantity ${\cal Q}$ is always positive and smaller than $\openone$, i.e. 
 \begin{eqnarray}
 \openone \geqslant {\cal Q} \geqslant 0\;.
 \end{eqnarray} 
Both properties simply follow from the identity 
 \begin{eqnarray}
{\cal Q} = P( {\openone} - \sum_{\vec{j}} {p_{\vec{j}}}P_{\vec{j}} )P   =P\big[ \sum_{\vec{j}} \;{p_{\vec{j}}}\;  (  {\openone} - P_{\vec{j}}  )\big] P
\;,
\end{eqnarray}
and from the fact that $ \openone \geqslant {\openone} - P_{\vec{j}}  \geqslant 0$.
We also notice that  
  \begin{eqnarray}\openone \geqslant 
 W_1  \geqslant  W_0 \; \; 2^{- n (S(\rho) -\chi({\cal E}) + \delta)} \geqslant 0\;,
 \end{eqnarray}  
where the last inequality is obtained by observing that the typical eigenvalues  $\lambda_{\vec{k}}^{(\vec{j})}$ are lower bounded as in Eq.~(\ref{def1}). 
From the above expressions we can conclude that 
the quantity in the summation that appears  on the lhs of Eq.~(\ref{ff144rffdd}) is always smaller than one and that it  
is decreasing with $\ell$. An explicit proof of this fact is as follows 
\begin{eqnarray}
0 &\leqslant& \mbox{Tr} [ W_1  \; {\cal Q}^\ell ] = \mbox{Tr} [ \sqrt{W_1} \; {\cal Q}^{\frac{\ell-1}{2}} \; {\cal Q} \; {\cal Q}^{\frac{\ell-1}{2} }\; \sqrt{W_1} ]  \nonumber \\ &\leqslant& 
 \mbox{Tr} [ \sqrt{W_1} \; {\cal Q}^{\frac{\ell-1}{2}} \; \openone \; {\cal Q}^{\frac{\ell-1}{2} }\; \sqrt{W_1} ] = \mbox{Tr} [ W_1  \; {\cal Q}^{\ell-1} ] \;, \nonumber 
\end{eqnarray} 
where we used the fact that the square root of a non negative operator can be taken to be non negative too (for a more detailed characterization of $W_0$ see Appendix~\ref{appB}).
A further simplification of the bound can be obtained by replacing the terms in the summation of Eq.~(\ref{ff144rffdd}) with the smallest addendum. 
This yields 
\begin{eqnarray}\label{fbrief}
1- \langle{P_{err}}\rangle 
 \geqslant  \left| {A}\right|^2 \;,
 \end{eqnarray} 
 where, using the fact that  $\bar{\openone}^2 = \bar{\openone}=P$, we defined
\begin{eqnarray}\label{Exp}
{A}  & := & \mbox{Tr} [ W_1  {\cal Q}^{N-1}]  = \sum_{z=0}^{N-1} \tiny{ \left( \begin{array}{c} N-1 \\ z \end{array} \right)} \; (-1)^z\; 
f_z , \\
f_z &:= & 
\mbox{Tr} [ W_1  \; P\; \bar{W}_0^z ] \;.
 \end{eqnarray} 
 It turns out that  the quantities $f_z$ defined above 
  are positive,  smaller than one, and decreasing in $z$. Indeed as shown in the Appendix~\ref{appa}
 they satisfy the inequalities  
\begin{eqnarray}\label{ffh}
0\leqslant f_z \leqslant f_0 \;\; 2^{-nz (\chi({\cal E}) - 2\delta )}
\qquad \mbox{for all  integer $z$,} 
\end{eqnarray} 
and, for each given $\epsilon$,  there exists a  sufficiently large $n_0$ such that for $n\geqslant n_0$ 
\begin{eqnarray} 
1-\epsilon \leqslant &f_0&  \leqslant 1 \;. \label{hh}
\end{eqnarray}
Using these expressions, we can derive the following bound on $A$, i.e. 
\begin{eqnarray}
A &=& 
f_0+  \sum_{z=1}^{N-1} \; \tiny{\left( \begin{array}{c} N-1 \\ z \end{array} \right) }  (-1)^z f_z\nonumber \\ 
&\geqslant&  f_0 -   \sum_{z=1}^{N-1} \; \tiny{\left( \begin{array}{c} N-1 \\ z \end{array} \right)} f_z  = 2 f_0 - \sum_{z=0}^{N-1} \;
\tiny{ \left( \begin{array}{c} N-1 \\ z \end{array} \right)} f_z \nonumber \\
&\geqslant&  2 f_0 -f_0  \sum_{z=0}^{N-1} \; \tiny{\left( \begin{array}{c} N-1 \\ z \end{array} \right)} 2^{-n z (\chi({\cal E}) - 2\delta )} \nonumber \\
&=&  f_0 \;[2 - (1 + 2^{-n (\chi({\cal E}) - 2\delta )})^{N-1}  ] 
\label{jj}\;,
\end{eqnarray} 
where in the first inequality we get a bound by taking all the terms
of $k \geqslant 1$ with the negative sign, the second from
(\ref{ffh}).
 Now, on one
hand if $N$ is too large the quantity on the rhs side will become
negative as we are taking the $N$ power of a quantity which is larger
than $1$.  On the other hand, if $N$ is small then for large $n$ the
quantity in the square parenthesis will approach $1$. This implies
that there must be an optimal choice for $N$ in order to have $ [2 -
(1 + 2^{-n (\chi({\cal E}) - 2\delta )})^{N-1} ]$ approaching one for
large $n$.  To study such threshold we rewrite Eq.~(\ref{jj}) as
\begin{eqnarray}
A  \geqslant  f_0
  [ 2 -Y(x=2^{\chi({\cal E}) - 2\delta }, y=N, n) ]\;,
 \label{jj22}
\end{eqnarray} 
where we defined 
\begin{eqnarray}
Y(x,y,n): = (1 + x^{-n})^{y^n-1}   \;.
\end{eqnarray}
We notice that for $x,y\geqslant  1$, in the limit of $n\rightarrow \infty$ the quantity $\log[Y(x,y,n)]$ 
is an indeterminate form. Its behavior can be studied for instance
using the de l'H\^{o}pital formula, yielding
 \begin{eqnarray}
\lim_{n\rightarrow \infty} \log[Y(x,y,n)]
= \frac{\log x}{\log y} \;  \lim_{n\rightarrow \infty} 
\; \left(\frac{y}{x}\right)^n 
 \;. 
\end{eqnarray}
This shows that if $y<x$ the limit exists and it is zero, i.e. $\lim_{n \rightarrow \infty} Y(x,y,n) =1$. Vice-versa for $y>x$ the limit diverges, and thus $\lim_{n \rightarrow \infty} Y(x,y,n) =\infty$.
Therefore, assuming 
 $N = 2^{nR}$, 
we can  conclude that as long as  
\begin{eqnarray}
R <  \chi({\cal E}) - 2\delta\label{condition2}\;,
\end{eqnarray} 
the quantity on the rhs of Eq.~(\ref{jj22}) approaches $f_0$ as $n$ increases (this corresponds to having $y<x$ in the $Y$ function).  
Reminding then Eq.~(\ref{hh}) we get
\begin{eqnarray}\label{f1brief}
1- \langle{P_{err}}\rangle 
 \geqslant  \left| {A}\right|^2 > f_0^2 > |1-\epsilon|^2 > 1 -2 \epsilon \;,
 \end{eqnarray} 
and thus
 \begin{eqnarray}\label{fbrief2}
\langle{P_{err}}\rangle 
< 2 \epsilon \;.
 \end{eqnarray} 
 On the contrary, if 
$R> \chi({\cal E}) - 2\delta $, the lower bound on $A$ becomes infinitely negative and hence useless to set a proper upper bound on $\langle{P_{err}}\rangle$. 

To summarize, we have shown that adopting the sequential detection strategy defined in Sec.~\ref{sec2} we can conclude that 
it is possible to send $N= 2^{nR}$ messages with asymptotically vanishing error probability, for all rates $R$ which satisfy the condition~(\ref{condition2}). 
  $\blacksquare$

\section{Conclusions}\label{seccon} 

To summarize: the above analysis provides an explicit upper bound for
the averaged error probability of the new detection scheme (the
average being performed over all codewords of a given code, and over
all possible codes). Specifically, it shows that the error probability
can be bound close to zero for codes generated by sources ${\cal E}$
which have strictly less than $2^{n \chi({\cal E})}$ elements. In
other words, our new detection scheme provides an alternative
demonstration of the achievability of the Holevo bound~\cite{HOL}.

An interesting open question is to extend the technique presented here
to a decoding procedure that can achieve the quantum capacity of a
channel \cite{sl,sh,dev,hay}.

\section*{Acknowledgments}
VG is grateful to P. Hayden, A. S. Holevo, K. Matsumoto, J. Tyson and
A. Winter for comments and discussions.

VG acknowledges support from the FIRB-IDEAS project under the contract
RBID08B3FM and support of Institut Mittag-Leffler (Stockholm), where
he was visiting while part of this work was done.  SL was supported by
the WM Keck Foundation, DARPA, NSF, and NEC. LM was supported by the
WM Keck Foundation.

\appendix
\section{Derivation of the POVM}\label{APOVM}
Here we provide an explicit derivation of the POVM~(\ref{ffd3}) associated with our iterative measurement procedure. 
It is useful to describe the whole process as a global unitary
transformation that coherently transfers the
information from the codewords to some external memory register.  

Consider, for instance, the first step of the detection scheme where Bob tries to determine whether  or not a given state $|\Psi\rangle \in {\cal H}^{\otimes n}$ corresponds
to the first codeword $\rho_{\vec{j}_1}$ of his list. 
The corresponding measurement can be described as the following  (two-step)  unitary transformation
\begin{eqnarray}
|\Psi\rangle |00\rangle_{B_1}  &\rightarrow&   P |\Psi\rangle |01  \rangle_{B_1} +   (\openone -P)  |\Psi\rangle |00 \rangle_{B_1} \nonumber \\
&\rightarrow& P_{\vec{j}_1} P |\Psi\rangle |1 1 \rangle_{B_1} +   (\openone -P_{\vec{j}_1}) P |\Psi\rangle |0 1 \rangle_{B_1} \nonumber \\
&&\qquad \qquad \qquad  \;  + \;  (\openone -P)  |\Psi\rangle |0 0 \rangle_{B_1} , \label{twostep}
\end{eqnarray} 
where $B_1$ represents a two-qubit memory register which stores the information extracted from the system. Specifically,  the first qubit 
records with a  ``1'' if the state $|\Psi\rangle$ belongs to the typical subspace ${\cal H}_{typ}^{(n)}$  of the average state of the source (instead it will
keep the value  ``0'' if this is not the case).
Similarly, the second qubit of $B_1$  records with a  ``1'' if  the projected component $P |\Psi\rangle$ is in the typical subspace  ${\cal H}_{typ}^{(n)}(\vec{j}_1)$ of 
$\rho_{\vec{j}_1}$. Accordingly the joint probability of success of finding  $|\Psi\rangle$ in ${\cal H}_{typ}^{(n)}$ and {\em then}  in    ${\cal H}_{typ}^{(n)}(\vec{j}_1)$
is given by
\begin{eqnarray}
{\cal P}_1(\Psi) = \langle \Psi | P  P_{\vec{j}_1} P |\Psi\rangle \;,\label{qui}
\end{eqnarray} 
in agreement with the definition of $E_1$ given in Eq.~(\ref{E1}).
Vice-versa the joint probability of finding the state $|\Psi\rangle$
in in ${\cal H}_{typ}^{(n)}$ and {\em then not} in ${\cal
  H}_{typ}^{(n)}(\vec{j}_1)$ is given by $\langle \Psi | P (\openone-
P_{\vec{j}_1}) P |\Psi\rangle$ and finally the joint probability of
{\em not} finding $|\Psi\rangle$ in in ${\cal H}_{typ}^{(n)}$ is
$\langle \Psi | \openone- P|\Psi\rangle$.  Let us now consider the
second step of the protocol where Bob checks wether or not the message
is in the typical subspace ${\cal H}_{typ}^{(n)}(\vec{j}_2)$ of
$\rho_{\vec{j}_2}$. It can be described as a unitary gate along the
same lines of Eq.~(\ref{twostep}) with $P_{\vec{j}_1}$ replaced by
$P_{\vec{j}_2}$, and $B_1$ with a new two-qubit register $B_2$. Notice
however that this gate only acts on that part of the global system
which emerges from the first measurement with $B_1$ in $|01\rangle$.
This implies the following global unitary transformation,
\begin{eqnarray}
&&|\Psi\rangle |00\rangle_{B_1}  |00\rangle_{B_2} \rightarrow
P_{\vec{j}_1} P |\Psi\rangle |1 1 \rangle_{B_1}  |00\rangle_{B_2} \nonumber \\
&&\qquad\qquad + \big[ \;   P_{\vec{j}_2} P (\openone -P_{\vec{j}_1}) P |\Psi\rangle |0 1 \rangle_{B_1} |11\rangle_{B_2} \nonumber \\ 
&&\qquad\qquad +   (\openone -P_{\vec{j}_2})  P (\openone -P_{\vec{j}_1}) P |\Psi\rangle |0 1 \rangle_{B_1} |01\rangle_{B_2} \nonumber \\ 
&&\qquad\qquad + (\openone - P)  (\openone -P_{\vec{j}_1}) P |\Psi\rangle |0 1 \rangle_{B_1} |00\rangle_{B_2} \; \big] \nonumber \\
&&\qquad\qquad +  (\openone -P)  |\Psi\rangle |0 0 \rangle_{B_1} |00\rangle_{B_2}  , \label{twostepb}
\end{eqnarray} 
which shows that the joint probability of finding $|\Psi\rangle$ in ${\cal H}_{typ}^{(n)}(\vec{j}_2)$ (after having found it in ${\cal H}_{typ}^{(n)}$, not in ${\cal H}_{typ}^{(n)}(\vec{j}_1)$, and
again in ${\cal H}_{typ}^{(n)}$) is 
\begin{eqnarray}
{\cal P}_2(\Psi) = \langle \Psi | P  (\openone -P_{\vec{j}_1}) P   P_{\vec{j}_2} P (\openone -P_{\vec{j}_1}) P |\Psi\rangle \;,\label{qua}
\end{eqnarray} 
in agreement with the definition of $E_2$ given in Eq.~(\ref{gg}).
Reiterating this procedure for all the remaining steps one can then
verify the validity of Eq.~(\ref{ffd3}) for all $u\geqslant 2$.
Moreover, it is clear (e.g. from Eq.~\eqref{qui} and \eqref{qua}) that it is
a quite different POVM from the conventionally used pretty good
measurement~\cite{HOL,SCHU}.

\section{Some useful identities}\label{appB}
In this section we derive a couple of inequalities which are not used in the main derivation but which allows us to better characterize the various operators which enter
into our analysis. 
First of all we observe that
\begin{eqnarray}\label{prima}
W_0 = \sum_{\vec{j}} p_{\vec{j}} P_{\vec{j}}  \leqslant \rho^{\otimes n} \;2^{n(S(\rho) - \chi({\cal E}) + \delta)}\;,
\end{eqnarray} 
which follows by the following chain of inequalities,
\begin{eqnarray}
W_0 &=& \sum_{\vec{j}} p_{\vec{j}} P_{\vec{j}} \nonumber 
=\sum_{\vec{j}} p_{\vec{j}} \; 
\sum_{\vec{k}\in{\cal K}_{\vec{j}} }
|e_{\vec{k}}^{(\vec{j})} \rangle \langle e_{\vec{k}}^{(\vec{j})} | \\
&\leqslant& \sum_{\vec{j}} p_{\vec{j}} \; 
\sum_{\vec{k}\in{\cal K}_{\vec{j}} }
|e_{\vec{k}}^{(\vec{j})} \rangle \langle e_{\vec{k}}^{(\vec{j})} | \lambda_{\vec{k}}^{(\vec{j})} \;2^{n(S(\rho) - \chi({\cal E}) + \delta)} \nonumber 
\\
&\leqslant& \sum_{\vec{j}} p_{\vec{j}} \; 
\sum_{\vec{k}}
|e_{\vec{k}}^{(\vec{j})} \rangle \langle e_{\vec{k}}^{(\vec{j})} | \lambda_{\vec{k}}^{(\vec{j})} \;2^{n(S(\rho) - \chi({\cal E}) + \delta)} 
\nonumber  \\ 
&= &  \sum_{\vec{j}} p_{\vec{j}} \; 
\rho_{\vec{j}}  \;2^{n(S(\rho) - \chi({\cal E}) + \delta)} \nonumber \\
& = &\rho^{\otimes n} \;2^{n(S(\rho) - \chi({\cal E}) + \delta)}\;, \nonumber 
\end{eqnarray} 
where we used Eq.~(\ref{def1}). 
 We can also prove the following identity
\begin{eqnarray}
{\cal Q} &=&  \sum_{\vec{j}} \; p_{\vec{j}}  \; \bar{Q}_{\vec{j}_\ell} = P(  {\openone} - {W}_0)P\nonumber \\  &\geqslant& 
 P(  {\openone} -  \rho^{\otimes n} \;2^{n(S(\rho) - \chi({\cal E}) + \delta)})P  \nonumber  \\
 &\geqslant& P \; (1 - 2^{-n( \chi({\cal E}) -2 \delta)})\;,
\end{eqnarray}
which follows by using Eq.~(\ref{use}). 
 Notice that due to Eq.~(\ref{prima})  this also gives
 \begin{eqnarray}\label{prima1}
PW_0 P \leqslant  P \;  2^{-n( \chi({\cal E}) -2 \delta)} \;.
\end{eqnarray} 
\newline
 
 \section{Characterization of the function$f_z$}\label{appa}

We start deriving the inequalities of Eq.~(\ref{hh}) first. 
To do we  observe that for all $\epsilon'$ positive we can write 
  \begin{eqnarray} \label{IMPO1}
\sum_{\vec{j}} \; p_{\vec{j}} \; \mbox{Tr} [\rho_{\vec{j}}  (\openone - P_{\vec{j}})P ] \leqslant 
\sum_{\vec{j}} \; p_{\vec{j}} \; \mbox{Tr} [ \rho_{\vec{j}} (\openone - P_{\vec{j}})] < 
 \epsilon' \;, \nonumber 
\end{eqnarray} 
  where the first inequality follows by simply noticing that $\rho_{\vec{j}}  (\openone - P_{\vec{j}})$ is positive semidefinite (the two operators commute), while the last is just Eq.~(\ref{IMPO})
  which holds for sufficiently large $n$. 
  Reorganizing the  terms and using Eq.~(\ref{bbhd1}) this finally yields 
    \begin{eqnarray} \label{IMPO2}
f_0 =  \mbox{Tr} [ W_1  \; P  ]  &>&  \sum_{\vec{j}} \; p_{\vec{j}} \; \mbox{Tr} [\rho_{\vec{j}} P ] -  
  \epsilon' \nonumber \\  &=& \mbox{Tr} [\rho^{\otimes n} P ]  -\epsilon' > 1 -2 \epsilon' \;,
\end{eqnarray} 
which corresponds to  the lefttmost inequality of  Eq.~(\ref{hh}) by setting  $\epsilon=2\epsilon'$. The rightmost inequality instead follows simply by observing that 
\begin{eqnarray}
 f_0 =  \mbox{Tr} [ W_1  \; P ] \leqslant   \mbox{Tr} [ W_1   ] = \sum_{\vec{j}} p_{\vec{j}} \mbox{Tr} [ P_{\vec{j}} \rho_{\vec{j}}] \leqslant 1\;.
  \end{eqnarray} 
  
To prove the inequality (\ref{ffh}) we finally notice that  for $z\geqslant 1$ we can write  
\begin{eqnarray} 
f_z &=& \mbox{Tr} [ W_1  \;  P \bar{W}_0^z]  = 
 \mbox{Tr} [ W_1  \; \bar{W}_0^z ]  \nonumber \\
 &= & \mbox{Tr} [ \sqrt{W_1 } \; \bar{W}_0^{\frac{z-1}{2} } \bar{W_0} \bar{W}_0^{\frac{z-1}{2} } \; \sqrt{W_1}]   \nonumber \\
 & \leqslant & \mbox{Tr} [ \sqrt{W_1 } \; 
 \bar{W}_0^{\frac{z-1}{2} }
 P\;  \bar{W}_0^{\frac{z-1}{2} } 
 \sqrt{W_1}]  
2^{-n(\chi({\cal E}) - 2\delta )} \nonumber \\ &\leqslant& \mbox{Tr} [ \sqrt{W_1 } \; 
 \bar{W}_0^{\frac{z-1}{2} }
  \bar{W}_0^{\frac{z-1}{2} } 
 \sqrt{W_1}]  
2^{-n(\chi({\cal E}) - 2\delta )} \nonumber \\
&=& \mbox{Tr} [W_1\bar{W}_0^{z-1} 
  ]\; 2^{-n(\chi({\cal E}) - 2\delta )}  = f_{z-1} \;2^{-n(\chi({\cal E}) - 2\delta )}  \nonumber \;,\end{eqnarray} 
where we used the fact that the operators operators $W_1$, $\bar{W}_0$  are non negative. 
The expression  (\ref{ffh}) then follows by simply reiterating the above inequality $z$ times.

\end{document}